\documentclass[12pt,preprint]{aastex}

\shorttitle{Cometary activity at 25.7 AU}
\shortauthors{Szab\'o et al.}

\begin{document}
\def\mycolumnwidth{8.0cm}
\def\myhalfcolumnwidth{3.84cm}
\def\myseventhcolumnwidth{1.04cm}
\def\myttcolumnwidth{5.20cm}
\def\mytncolumnwidth{0.32cm}

\title{Cometary activity at 25.7 AU: Hale--Bopp 11 years after perihelion\altaffilmark{1}}

\author{Gy. M. Szab\'o\altaffilmark{2,3,4},
L. L. Kiss\altaffilmark{2},
K. S\'arneczky\altaffilmark{2,5,6}
}
\altaffiltext{1}{Based on data obtained at the Siding Spring  Observatory}
\altaffiltext{2}{School of Physics, University of Sydney, NSW 2006, Australia}
\altaffiltext{3}{Department of Experimental Physics \& Astronomical Observatory,
      University of Szeged, 6720 Szeged, Hungary} 
\altaffiltext{4}{Group of Eight European Fellow; szgy@titan.physx.u-szeged.hu}
\altaffiltext{5}{Department of Optics and Quantum Electronics, University of Szeged, 6720 Szeged, Hungary}
\altaffiltext{6}{Hungarian State E\"otv\"os Fellow}

\begin{abstract}
Eleven years after its perihelion, comet C/1995 O1 (Hale-Bopp) is still active. Between 2007 October 20 and 22, we detected a diffuse coma of 180$\times 10^3$ km in diameter with a slight elongation toward the north-south direction. 
The integrated brightness was 20\fm04 in $R_C$, implying $Af\rho=300$ m and albedo$\times$ dust surface $a_RC=4300$ km$^2$. The coma was relatively red at $V-R=0.66$ mag, which is consistent with that of the dust in other comets. The observed properties and the overall fading in brightness between 10 AU and 26 AU follow the predicted behaviour of CO-driven activity. This is the most distant cometary activity ever observed.

\end{abstract}


\section{Introduction.}

Long-period comets experience little solar exposure and heating, which means they can be used to reveal conditions that existed during the formation and early evolution of the solar system (Delsemme 1977; Lowry \&{} Fitzsimmons 2005). The main evidence supporting this argument is that the post-perihelion activity decreases fast after the cut-off of water sublimation at 3 AU (Fern\'andez 2005), and then the activity usually stops until the next apparition. A few exceptional comets, however, displayed activity far beyond 3 AU, which can be explained by the sublimation of CO, but the processes involved are not well understood (Mazzotta Epifani et al. 2007). Because of this, cometary activity at large heliocentric distances has raised a considerable interest recently, addressing the question of how intact the matter in comets is (Lowry et al. 1999). For example, dust activity throughout the entire orbit could result in a continuous resurfacing of the nucleus, while the surface composition could also be altered. The observed diversity of nucleus colors (Luu 1993; Jewitt 2002) may indicate that this is indeed the case. On the other hand, comets with disguised distant activity will have larger mass loss rate per orbit than we estimate, leading to overestimated comet lifetimes (Lowry et al. 1999; Mazzotta Epifani et al. 2007) and underestimated replenishment rate of the zodiacal dust (Liou et al. 1995).

In recent years, a number of studies reported on short-period comets active between 3 and 7 AU (e.g. Lowry et al. 1999; Lowry \&{} Fitzsimmons 2001, 2005; Lowry \&{} Wiessman 2003; Snodgrass et al. 2006, 2008; Mazzotta Epifani et al. 2006, 2007). These surveys aimed at the investigation of the bared nucleus, but a surprisingly high number of comets showed comae and even dust tails at large ($R>3)$ heliocentric distances, where volatile sublimation is expected to be low. The activity of some long period comets is similar, with ocassionally long dust tails (Szab\'o et al. 2001, 2002), while 11 Centaur objects are also known with cometary activity (see e.g. Rousselot 2008 and references therein). Chiron is known to be active at solar distances between 8 and 14 AU (Meech et al. 1997), and was seen to display considerable outgassing near aphelion (at 17.8--18.8 AU) between 1969 and 1977 (Bus et al. 2001). Meech et al. (2004) found that the Oort-cloud comet C/1987 H1 (Shoemaker) displayed an extensive tail at all distances between 5 and 18 AU, which is, as of this writing, the most distant example of cometary activity.

Discovered at 7.2 AU from the Sun, C/1995 O1 (Hale--Bopp) has been a prime target for cometary studies. In prediscovery images it had a faint coma 0\farcm4 in diameter and a total magnitude of $\sim$18 (McNaught \&{} Cass 1995), while the dust production rate was $\approx 500$ kg/s (Fulle et al. 1998) at a solar distance of 13.1 AU. At 7.0 AU, NIR absorption of water ice was detected (Davies et al. 1997). At that time the activity was driven by CO production (Biver et al. 1996, Jewitt et al. 1996), which switched to a water-driven activity at around 3 AU (Biver et al. 1997, Weaver et al. 1997). Approaching the 0.9 AU perihelion distance, the dust production rate was $2\times 10^6$ kg/s (Jewitt \&{} Matthews, 1999). The size distribution of the dust, especially in the jets, showed a dominance of $\lesssim0.5\ \mu$m grains, smaller than in any other comets. This was indicated by the unprecedentedly large superheat ($T_c / T_{bb}$ between 1.5--2 [Hayward et al. 2000] or 1.5--1.8 [Gr\"un et al. 2001]), and the scattering albedo and polarization (Hayward et al. 2000). The water production was $~10^{31}$ molecule/s, the largest value ever observed. The dust to gas ratio was very high, between 5 and 10, regardless of the solar distance (Colom et al. 1997, Lisse et al. 1997, Weaver et al. 1999). The production rates observed post-perihelion were similar to those observed pre-perihelion (Capria et al. 2002), foreshadowing long-lasting distant activity. We have indeed detected evidence for cometary activity at 25.7 AU from the Sun; this Letter presents our major findings based on broadband imaging in late 2007.

\section{Observations}

New observations were taken with the 2.3 m ANU telescope at the Siding Spring Observatory on 2007 October 20, 21 and 22. The solar distance of the comet was 25.7 AU. We took 9$\times$240 s exposures in Johnson-Cousins $VR_C$ filters with a 2$\times$2 binned image scale of 0\farcs67/pixel. The seeing was 2\farcs0--2\farcs5 on the three nigths (see Table 1 for exposure data and ephemerides). 

\begin{figure}
\begin{center}
\includegraphics[width=\mycolumnwidth]{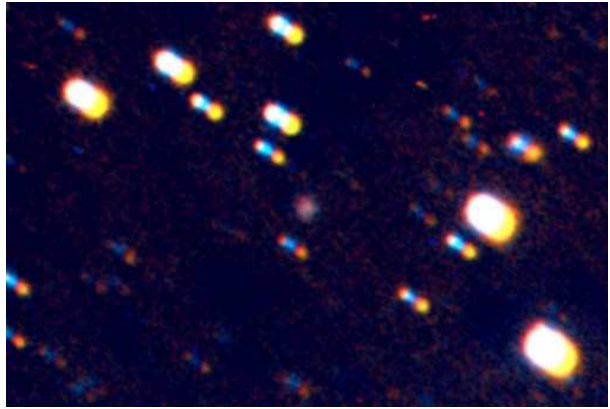}
\end{center}
\caption{Comet Hale--Bopp on October 21, 2007. The field of view is 5\farcm7$\times$3\farcm8; north is up, east is to the left.}
\label{zsaner}
\end{figure}

\begin{figure}
\begin{center}
Oct 20 \hskip3.5cm Oct 21\par
\includegraphics[height=\myhalfcolumnwidth]{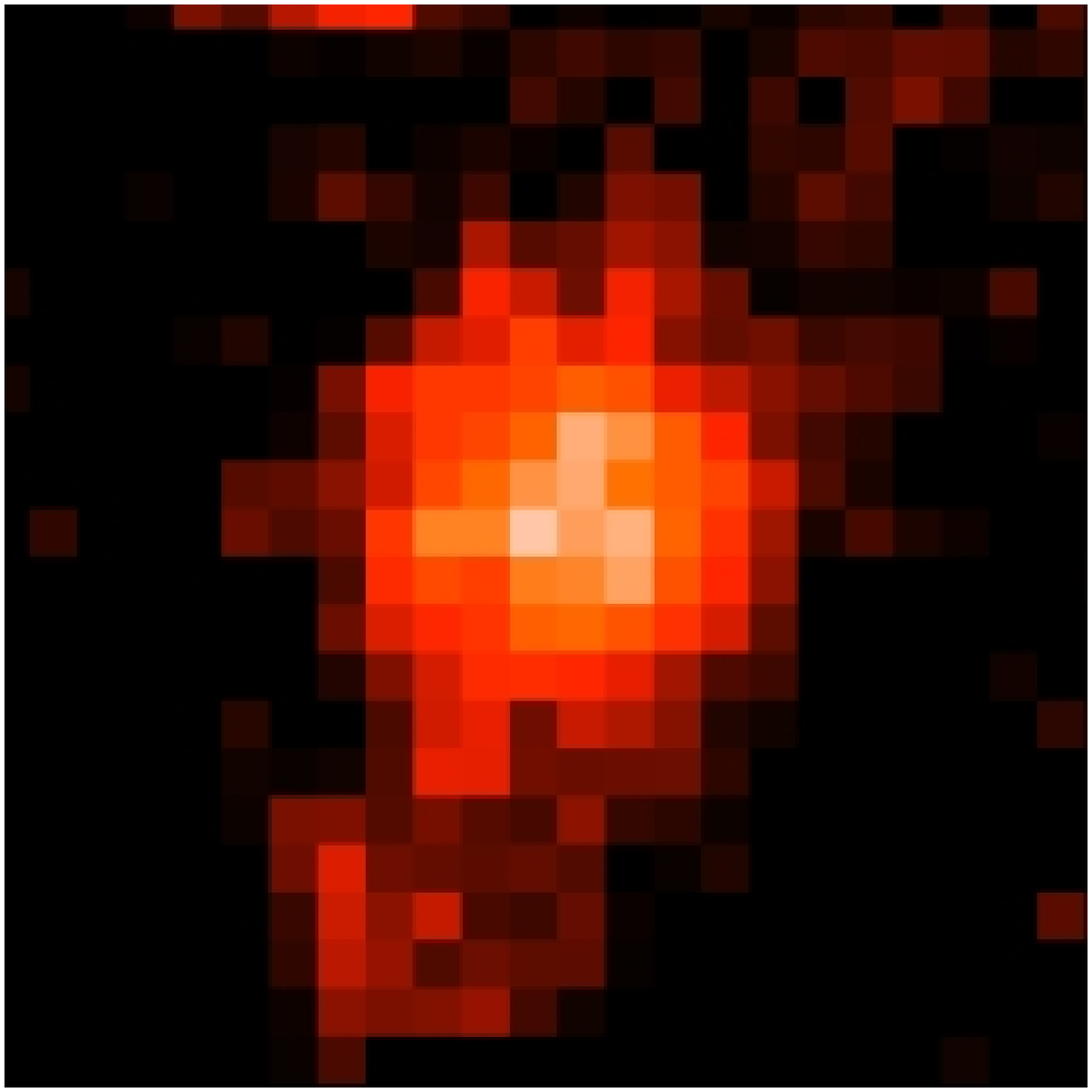}\hskip\mytncolumnwidth%
\includegraphics[height=\myhalfcolumnwidth]{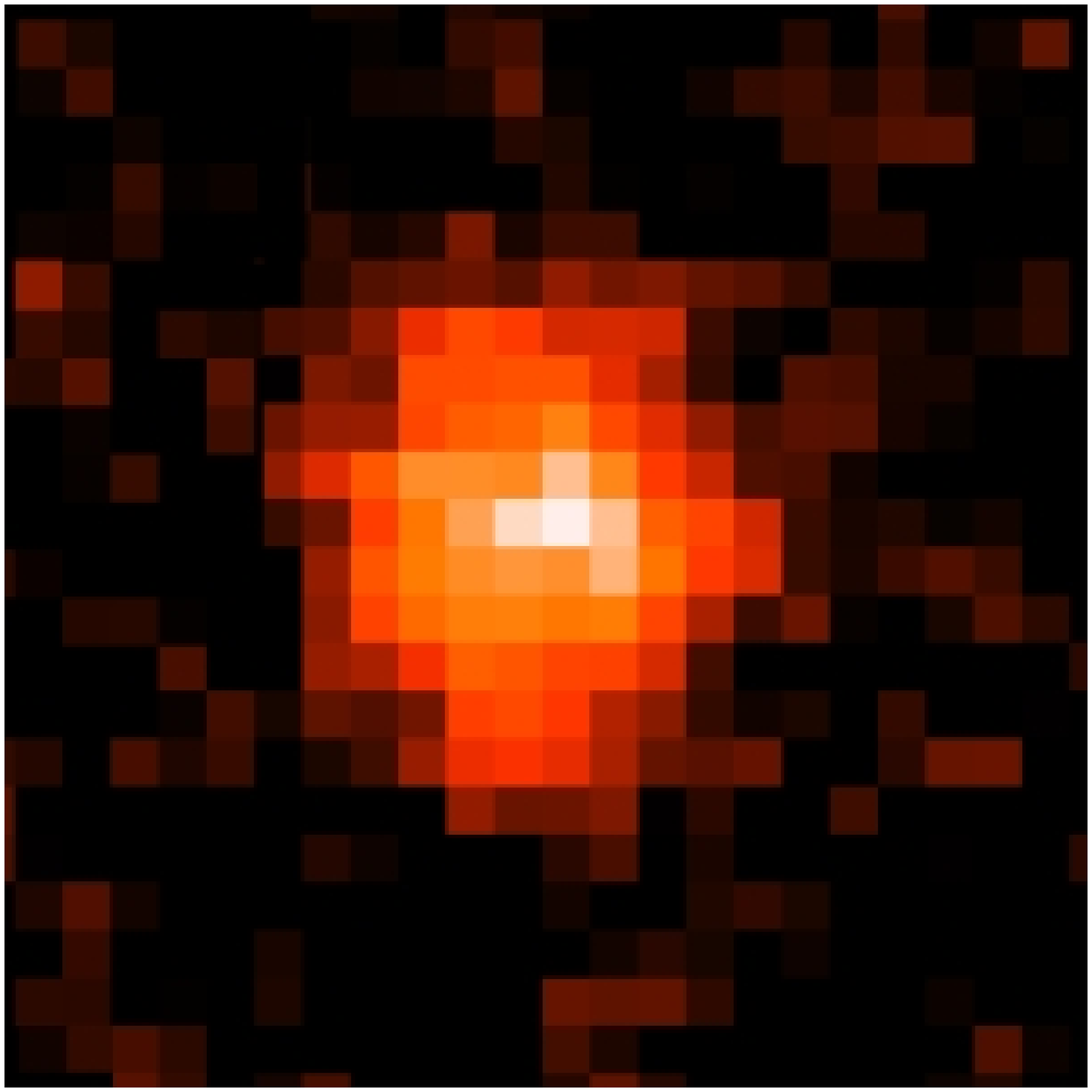}\par
Oct 22 \hskip3.5cm Oct 21, V\par
\includegraphics[height=\myhalfcolumnwidth]{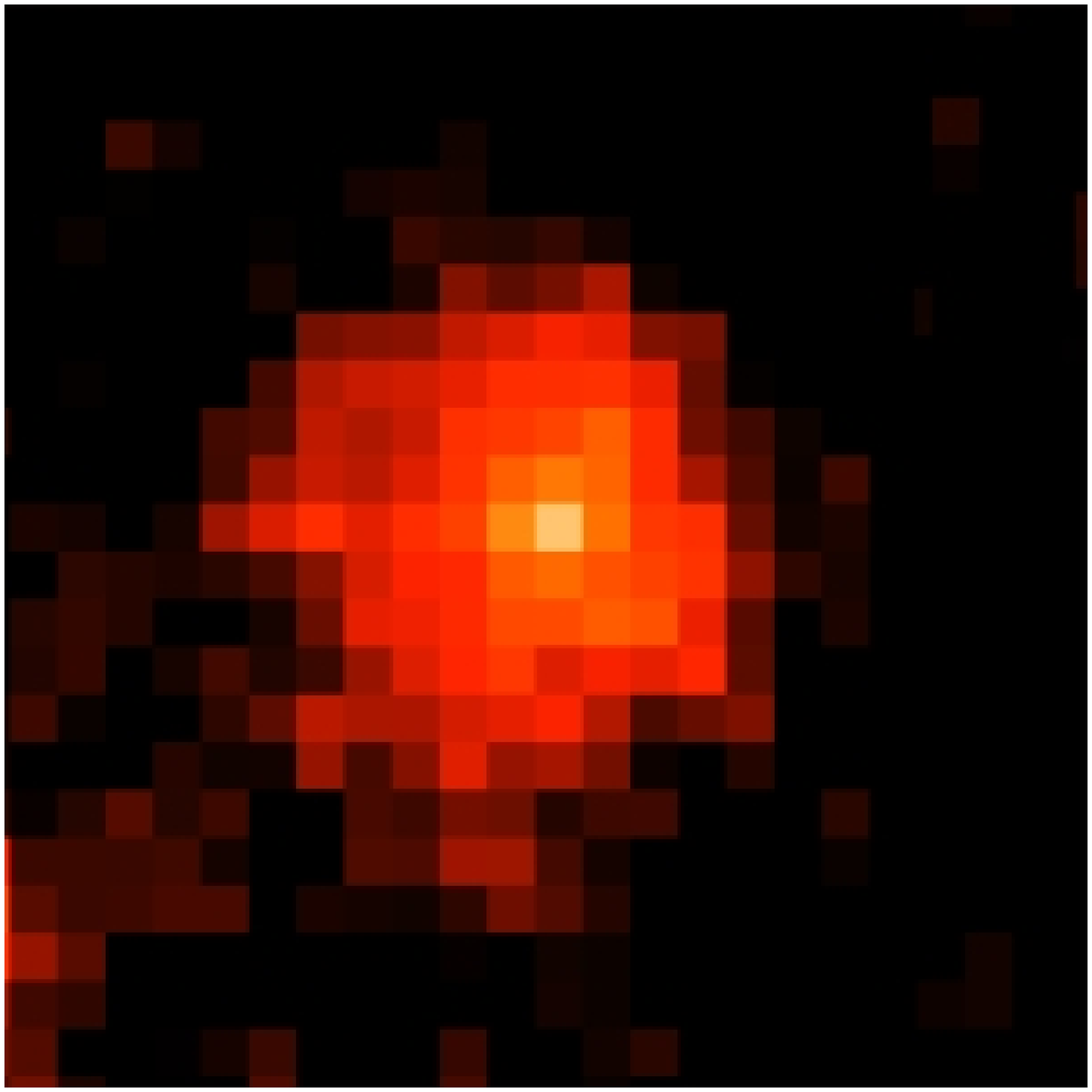}\hskip\mytncolumnwidth%
\includegraphics[height=\myhalfcolumnwidth]{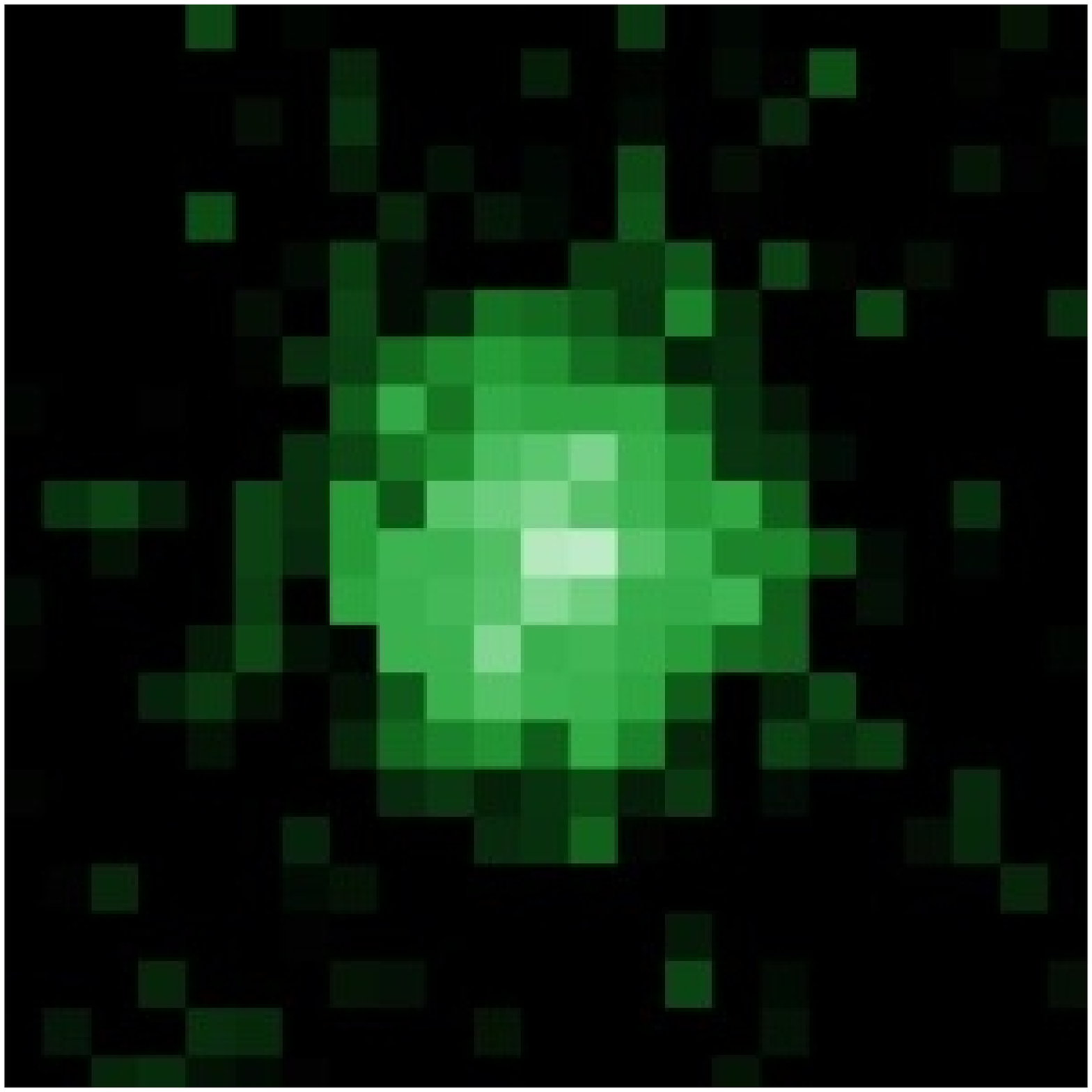}
\end{center}
\caption{The $R_C$ and $V$ images of comet Hale-Bopp between Oct 20--22, 2007. The field of view is $24\farcs1\times24\farcs1$ $(300\times 10^3$ km).}
\label{ME}
\end{figure}

The images were corrected in a standard fashion, including bias and flat-field correction and fringing correction of the $R_C$ images. Every night we aligned and co-added the images by fitting a coordinate grid to the stars, yielding a ``star field'' image for photometric calibrations. The images were then re-aligned with respect to the proper motion of the comet, to get untrailed ``comet'' images. In this step, the MPC ephemerides at the time of each observation were used to match the individual frames. Fig. \ref{zsaner} shows the ``comet'' image on 2007 October 21. The estimated size of the coma is 180$\times 10^3$ km, slightly elongated north/southward (Fig. \ref{ME}).

\subsection{Photometry}

A good proxy to the dust content inside the coma is its brightness, which we have measured by aperture photometry in a single aperture of 14\arcsec{} across. On October 21, the comet and field stars were calibrated with all-sky photometry using the SA 98 field  of Landolt (1992), observed between airmasses 1.17 and 1.65 (Hale--Bopp was at $X=1.72$). Due to the slow apparent motion, we could use the same field stars as local standards on the other two nights as well. The measured brightnesses on October 21 were $V=20\fm70\pm0.1$, $R_C=20\fm04\pm0.1$. This corresponds to $Af\rho=30~000$ cm, according to the definition by A'Hearn et al. (1984). For comparison, this value is twice as large as that for 29P/Schwassmann-Wachmann 1 in outburst (Szab\'o et al. 2002), and 3 times larger than for 174P/Echeclus in outburst (Rousselot, 2008).

\subsection{Morphology}

The observed brightness can be converted to albedo $\times$ dust surface, $a_RC$ (Eddington, 1910), which is the cross section of reflecting particles ($C$, in m$^2$) in the aperture, multiplied by the $a_R$ geometric albedo in the $R_C$ photometric band. It is calculated as 
\begin{equation}
a_RC = {2.22\times 10^{22} \pi R^2 \Delta^2 10^{0.4(m_{\sun} - m_{\rm comet})} \over 10^{-0.4\alpha\beta}},
\end{equation} 
where $m_{\sun}=-27\fm11$, the apparent $R_C$ brightness of the Sun, and the $\beta$ phase coefficient is usually assumed to be 0.04. Substituting the measured total brightness yields $a_RC\approx 4300$ km$^2$. 
For comparison, this cross section is 450 times larger than that of the dust cloud ejected in the Deep Impact experiment (e.g. Milani et al. 2007). After calibrating the image flux, the azimuthally averaged comet profile was determined from surface photometry. From this the local filling factor of the dust, $f$, can directly be expressed by replacing $m_{\rm comet}$ with the $\mu$ surface brightness:
\begin{equation}
a_Rf={1.34\times 10^{17} R^2 10^{0.4(m_{\sun} - \mu)} \over 10^{-0.4\alpha\beta}},
\end{equation}
that is the measured surface brightness relative to that of a reflecting surface with 1.0 albedo. We found that the surface brightness was 20\fm3 in the inner coma, corresponding to $a_Rf \approx 9\times 10^{-6}$, which remained above $10^{-6}$ in the inner 70~000 km (Fig. 3). 

\section{Discussion}

As we describe below, the observations are consistent with a CO-driven activity. Following Fern\'andez (2005), the thermal equilibrium of the absorbed radiation, the emitted radiation and the latent heat lost by sublimation can be written as 
\begin{equation}
{F_{\sun}\over R^2}\pi r^2 = 4 \pi r^2 \sigma T^4 + 4\pi r^2 f \zeta(T) l_s,
\end{equation} 
where $r$ is the radius of the nucleus, $T$ is the temperature, $f<1$ is the fraction of the active area, $\zeta(T)$ is the gas production rate in molecules/m$^2$/s, and $l_s$ is the latent heat loss per one molecule. CO molecules deposited on CO ice (e.g. inclusions) and CO molecules deposited on H$_2$O can sublimate at large solar distances, due to their high volatility. By neglecting the CO ice inclusions, CO is condensed on water ice, for which $l_s=10^{-20}$ J/molecule (Delsemme 1981), and $\log\zeta(T)\approx 755.7/T - 35.02$ (Mukai et al. 2001). The heat loss of such a nucleus is plotted in Fig. \ref{sublimfig} for different values of $f$ and as a function of temperature.

The equivalent temperature of a freely sublimating ice globe ($f=1$) at 25.7 AU is 48.0 K, which is slightly less than 54.8 K for a blackbody ($f=0$), due to the sublimation of $2\times 10^{19}$ molecules/m$^2$/s. This corresponds to $Q(CO)=4\pi r^2\zeta(T)=2.1\times 10^{20} r^2$ molecule/s. If the active area covers 1\%{} of the surface, the equilibrium temperature is 53.1 K, $\zeta(T)=6.2\times 10^{20}$ molecule/s/m$^2$, $Q(CO)=8\times 10^{19} r^2$ molecule/s. The $Q(CO)$ production rate and the temperature just slightly depend on $f$, thus we can get a reasonable estimate assuming $f=0.01$. The thermal velocity of the gas is $u_g=\sqrt{3kT/m_{\rm CO}} = 210$ m/s, which is enough to carry off small dust grains ($m_{\rm CO}$ is the mass of one CO molecule).
The acting drag force $F_D=\pi a^2 u_g \zeta(T) m_{\rm CO},$ where $a$ is the size of the dust particle. Particles are carried off if the drag force exceeds the gravitation $F_D>F_G=(4 \pi /3)^2{G\rho_n\rho_p a^3 r},$ $\rho_n$ and $\rho_p$ are the density of the nucleus and the dust particle, respectively. Thus, the maximum radius of the escaping dust particles is 
\begin{equation}
a_{\rm max} = {9\over 16 \pi} {u_g \zeta(T) m_{\rm CO} \over \rho_n\rho_p G r}.
\end{equation} 
For an order-of-magnitude estimate we assumed $\rho_n=1000$ kg/m$^3$ (Capria et al. 2002), $\rho_p=2500$ kg/m$^3$, $r=15$ km (Meech et al. 2004)  or $r=30$ km (Fern\'andez, 2000), leading to $a_{max}\approx 100\ \mu$m, and $Q(CO)\approx 1.7\times 10^{28}/$s$ = 790$ kg/s in the case of the $r=15$ km nucleus and $Q(CO)\approx 6.8\times 10^{28}/$s$ = 3160$ kg/s if $r=30$ km.

With a more sophisticated model (the surface is dominated by water ice instead of CO on water, the majority of CO is present in inclusions, crystallization heats the nucleus, the nucleus rotates), Capria et al. (2002) predicted $Q(CO)=5\times 10^{27}$ molecule/s (equivalent to $230$ kg/s) at 25 AU (see Fig. 4 in their paper). Assuming that the dust loading remained high ($m_{dust}/m_{gas}=$1--10), the total dust production ranges from 230 to 2300 kg/s. 
For example, if 500 kg/s dust is produced in the form of 1 $\mu$m-sized particles, the projected area is 0.25 km$^2$. Assuming a 0.05 albedo results in a 12,000 m$^2$ excess of $a_RC$ every second. With these assumptions, the nucleus can produce the observed amount of matter in the coma within $\sim$5 days. The measured radius of the coma and the time-scale of dust production needed gives a measure of dust ejection velocity as $90\ 000$~km$/5$ days$ \approx 210$~m/s, which is consistent with the thermal gas velocity derived at 25.7 AU.

This self-consistent picture is also supported by the brightness variation of Hale--Bopp between 10 AU and 26 AU. In Fig. \ref{lcs}, we plot the observed brightness of Hale--Bopp (collected from ICQ and MPC bulletins) against the solar distance. For comparison, data for 6 dynamically young Oort-comets are also plotted (Meech et al. 2004). Hale--Bopp was consistently 3--5 magnitudes brighter than these Oort-comets at all distances $\lesssim$15 AU. Beyond that, other comets quickly disappeared as opposed to Hale--Bopp, which kept its slow rate of fading. We also show a theoretical light curve predicted from the CO production curve by Capria et al. (2002), after scaling with $R^{-2}\Delta^{-2}$ and a constant dust loading. The overall agreement indicates that the distant brightness change is consistent with a CO-driven activity.

An alternative explanation of the present activity of Hale--Bopp could be that the light halo is not a real coma but a preserved debris tail (e.g. Jenniskens et al. 1997; Sekanina et al., 2001).
Comets with periods over a few thousand years are unlikely to preserve dense debris tail, but, as Lyytinen \& Jenniskens (2003) remarks, giant comets such as Hale--Bopp can be exception. At the current position, the orbital path of Hale--Bopp is almost parallel to the line of sight, with only 5$^\circ$ inclination and a solar phase angle of 2.2$^\circ$. Thus, a hypothetical thin, $\sim$1--2 million km long debris tail would appear as a $\sim$100--200 thousand km long ``tail'' in projection, which is approximately the diameter of the light halo around Hale--Bopp. However, this scenario  is not likely, because the optically detected dust trails are all very thin, only $\sim$10--20$\times 10^3$ km wide, and their projected images always appear as a thin feature pointing out of the nucleus (Ishuguro et al. 2007; Sarugaku et al. 2007). Our observation of a nearly spherical light halo, with the nucleus approximately in the center, is not compatible with such a dust trail.

\section{Conclusion}
Comet Hale--Bopp has been the single most significant comet encountered by modern astronomy and 11 years after perihelion it still displays fascinating phenomena. The detected activity can be well explained by theoretical models invoking CO sublimation at large heliocentric distances. Compared to other young Oort-cloud comets, the long-term behaviour of Hale--Bopp seems to suggest genuine differences beyond the difference in nucleus size.

The main results of this paper can be summarized as follows:
\begin{enumerate}
\item{} We detected cometary activity of Hale--Bopp at 25.7 AU, which is the most distant activity detection so far. 
\item{} Our analysis indicates that the extrapolation of the Capria et al. (2002) model works very well for the distant Hale--Bopp, which confirms the physical assumptions of this model.
\item{} Further observations with 8 m-class telescopes can help constrain the presence of gas in the coma, effects of superheat due to small dust particles, and, ultimately, the cessation of mass loss processes in Hale--Bopp.
\end{enumerate}

\acknowledgments This research was supported by the ``Bolyai J\'anos'' Research Fellowship of the Hungarian Academy of Sciences and a Group of Eight European Fellowship (Gy.M.Sz.), a University of Sydney Research Fellowship (L.L.K.) and a Hungarian State ``E\"otv\"os'' Fellowship (K.S.).

\clearpage

\begin{table*}
\caption{Ephemerides and exposure data}
\begin{tabular}{lllllllllllll}
\tiny Date (UT) &\tiny RA &\tiny Dec &\tiny $\lambda$ $[^\circ]$ &\tiny $\beta$ $[^\circ]$ &\tiny R [AU] &\tiny $\Delta$ [AU]&\tiny E $[^\circ]$ &\tiny $\alpha$ $[^\circ]$ &\tiny V exp (s) & \tiny S(\arcsec) &\tiny R exp (s) &\tiny S(\arcsec)\\
\hline
\tiny 2007-Oct-20  &\tiny   04 11 58.98 &\tiny $-$86 27 28.7 &\tiny 280.76 &\tiny $-$70.31& \tiny25.75 &\tiny 25.86  &\tiny 82.69 &\tiny  2.20 &---& --- &\tiny 9$\times$240  &\tiny 2.5 \\
\tiny 2007-Oct-21  &\tiny  04 09 56.22 &\tiny $-$86 28 46.5 &\tiny 280.76 &\tiny $-$70.31& \tiny 25.76 &\tiny 25.87  &\tiny 82.37 &\tiny 2.20 &\tiny 9$\times$240  &\tiny 2.0 &\tiny 9$\times$240  &\tiny 2.1 \\
\tiny 2007-Oct-22  &\tiny    04 07 53.57 &\tiny $-$86 30 04.2 &\tiny 280.76 &\tiny $-$70.31 &\tiny 25.77 &\tiny 25.88 &\tiny 81.98 &\tiny 2.20 &\tiny ---& ---& \tiny 9$\times$240  &\tiny 2.2\\
\hline
\end{tabular}
\end{table*}

\clearpage

\begin{figure}
\begin{center}
\includegraphics[width=\mycolumnwidth]{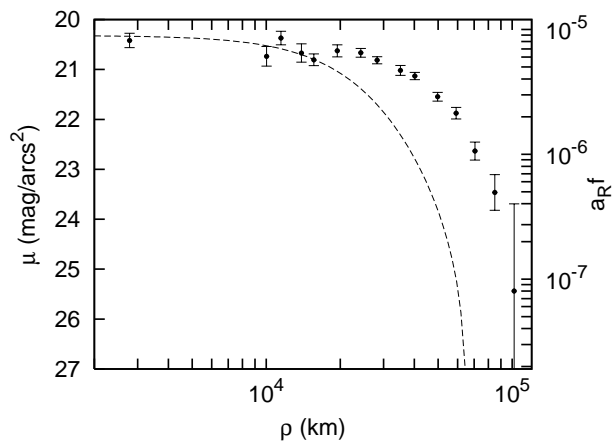}
\end{center}
\caption{The $R_C$ surface brightness profile of comet Hale-Bopp on October 21, 2007 with a stellar profile for comparison. The presence of the extended coma is obvious. The right axis displays $a_Rf$, i.e. the relative dust content per unit projected area inside the coma.}
\end{figure}

\begin{figure}
\begin{center}
\vskip0.1cm
\includegraphics[bb=108 79 374 185,width=\myttcolumnwidth]{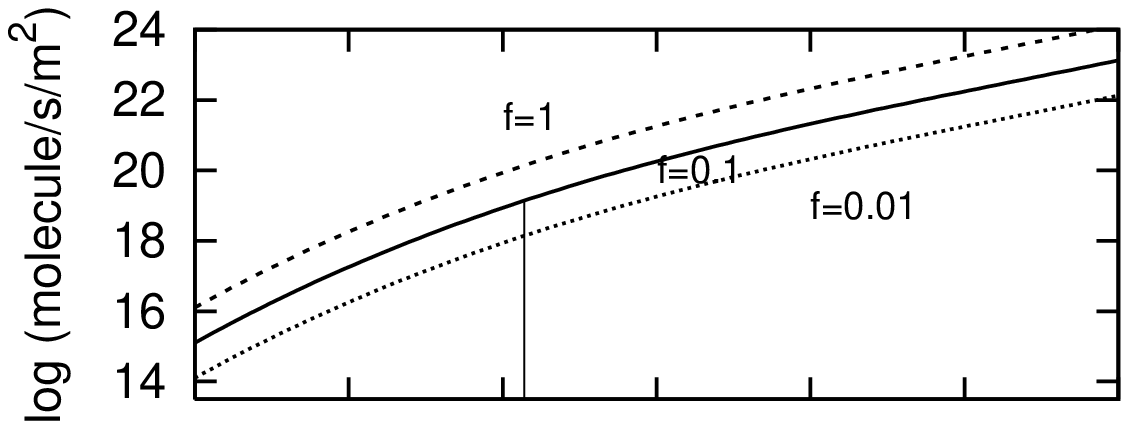}
\par
\vskip0.1cm
\includegraphics[bb=107 93 348 285,width=\myttcolumnwidth]{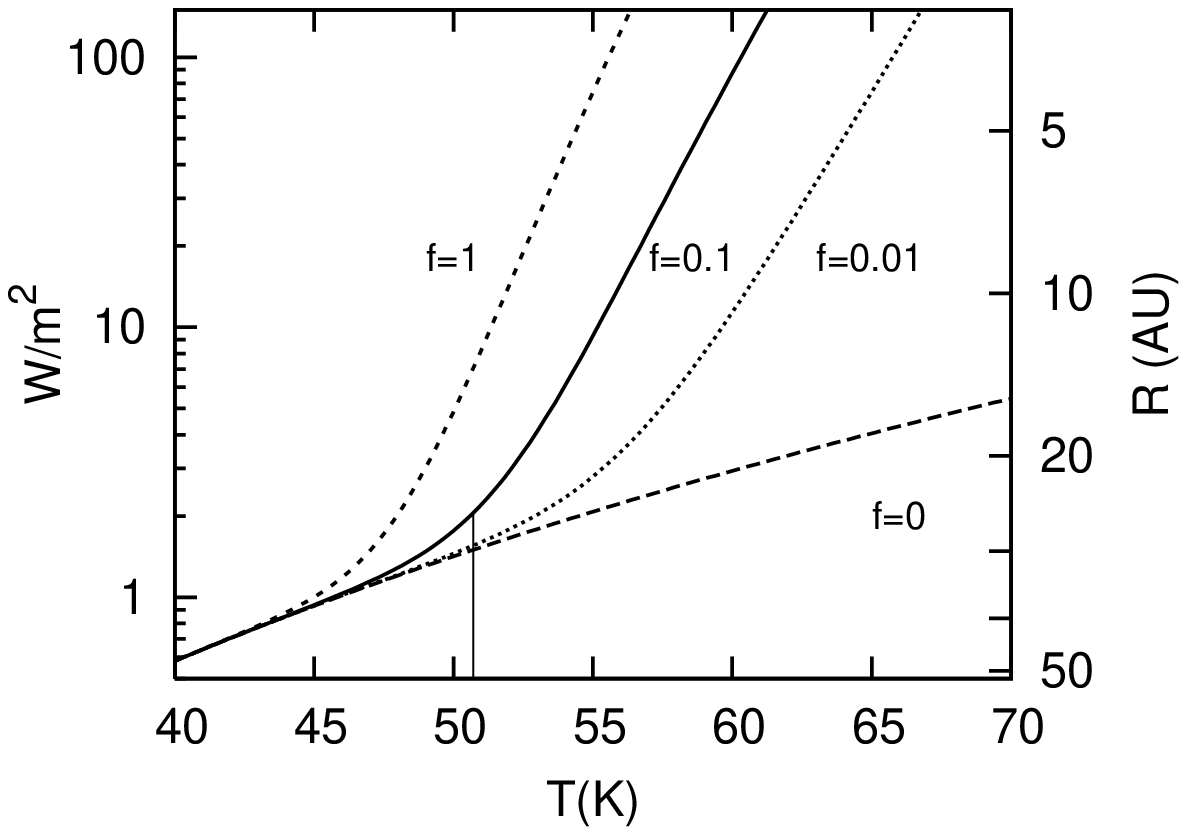}
\par\vskip1.0cm
\end{center}
\caption{Sublimation of CO depleted onto water ice. Top: the temperature dependence of sublimation rates (from Mukai et al. 2001). Bottom: the energy loss of radiation and sublimation for various $f$ fractions of active area. The right axis shows the distance scale that corresponds to the solar irradiation values on the left axis.}
\label{sublimfig}
\end{figure}

\begin{figure}
\begin{center}
\includegraphics[width=\mycolumnwidth]{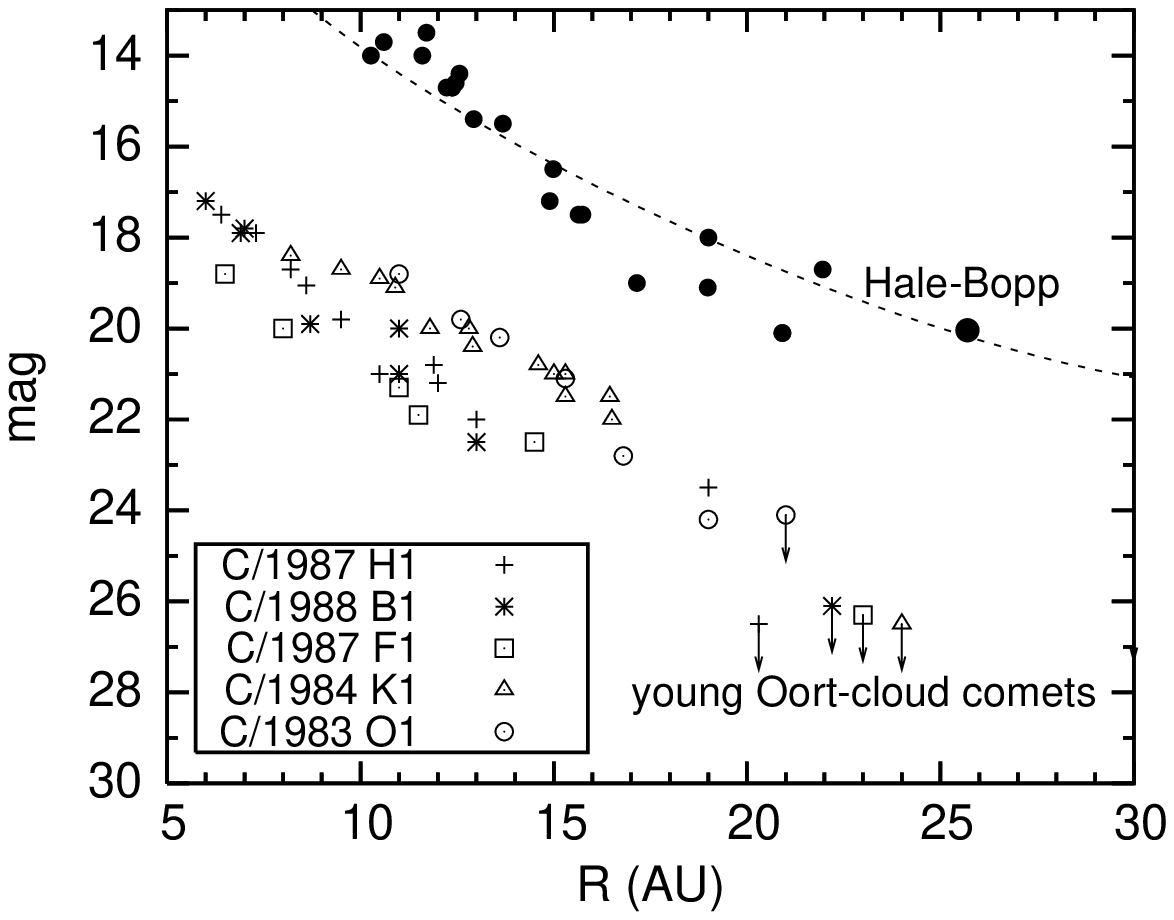}
\end{center}
\caption{The light curve of Hale--Bopp compared to observations of 6 dynamically young Oort-cloud comets (Meech et al. 2004; the last five point are upper limits). The dashed curve shows the prediction of the activity model (Capria et al. 2002).}
\label{lcs}
\end{figure}


\begin{thebibliography}{}

\bibitem[]{118} A'Hearn, M.F. et al., 1984, AJ, 89, 579

\bibitem[]{120} Biver, N. et al., 1996, Nature, 380, 137

\bibitem[]{122} Biver, N. et al., 1997, Science, 275, 1915

\bibitem[]{124} Bus, S.J., A'Hearn, M.F., Bowell, E., Stern, S.A., 2001, Icarus, 150, 94

\bibitem[]{126} Capria, M.T., Coradini, A., De Sanctis, M.C., 2002, EM{}P, 90, 217

\bibitem[]{128} Colom, P., G\'erard, E, Crovisier, J., Bockel\'ee-Morvan, D., Biver, N., Rauer, H., 1997, EM{}P, 78, 37

\bibitem[]{130} Davies, J.K. et al., 1997, Icarus, 127, 238 

\bibitem[]{132} Delsemme, A.H., 1977, in: Comets, Asteroids, Meteors, (ed. Delsemme, A.H.), Univ. Toledo, Ohio, p. 3

\bibitem[]{134} Delsemme, A.H., 1981, in: Comets (ed. Wilkening, L.L.), Univ. Arizona Press, Tucson, p. 85

\bibitem[]{136} Eddington, A.S., 1910, MNRAS, 70, 442

\bibitem[]{138} Fern\'andez, J.A., 2005, Comets, Astrophysics and Space Library, Springer, Dordrecht, NL
\bibitem[]{139} Fern{\'a}ndez, Y.R., 2000, EMP, 89, 3
\bibitem[]{140} Fulle, M., Cremonese, G., B\"o{}hm, C., 1998, AJ, 116, 1470

\bibitem[]{142} Gr\"un, E. et al., 2001, A\&{}A, 377, 1098

\bibitem[]{144} Hayward, T.L., Hanner, M.S., Sekanina, Z., 2000, ApJ, 538, 428

\bibitem[]{146} Ishiguro, M. et al., 2007, Icarus, 189, 169

\bibitem[]{148} Jenniskens, P., Betlem, H., de Lignie, L.M., 1997, ApJ, 479, 441

\bibitem[]{150} Jewitt, D., 2002, AJ, 123, 1039

\bibitem[]{152} Jewitt, D. et al., 1996, Science, 271, 1110

\bibitem[]{154} Jewitt, D., Matthews, H., 1999, AJ, 117, 1056

\bibitem[]{156} Landolt, A.U., 1992, AJ, 104, 340

\bibitem[]{158} Liou, J.C., Dermott S.F., Xu Y.L., 1995, P\&{}SS, 43, 717

\bibitem[]{160} Lisse, C.M. et al., 1997, EM{}P, 78, 251 

\bibitem[]{162} Lowry, S.C., Fitzsimmons, A., Cartwright I.M., Williams, I.P., 1999, A\&{}A, 349, 649

\bibitem[]{164} Lowry, S.C., Fitzsimmons, A., 2001, A\&{}A, 365, 204

\bibitem[]{166} Lowry, S.C., Weissman P.R., 2003, Icarus, 164, 492

\bibitem[]{168} Lowry, S.C., Fitzsimmons, A., 2005, MNRAS, 358, 641

\bibitem[]{170} Luu, J.X., 1993, Icarus, 104, 138

\bibitem[]{172} Lyytinen, E.,  Jenniskens, P., 2003, Icarus, 162, 443

\bibitem[]{174} Mazzotta Epifani, E., Palumbo, P., Capria, M.T.,  Cremonese, G., Fulle, M., Colangeli, L., 2006, A\&{}A, 460, 935

\bibitem[]{176} Mazzotta Epifani, E., Palumbo, P., Capria, M.T.,  Cremonese, G., Fulle, M., Colangeli, L., 2007, MNRAS, 381, 713

\bibitem[]{178} McNaught, R.H., Cass, C.P., 1995, IAU Circ., 6198

\bibitem[]{180} Meech, K.J., Buie, M.W., Samarasinha, N.H., Mueller, B.E.A., Belton, M.J.S., 1997, AJ, 113, 844

\bibitem[]{182} Meech, K.J., Hainaut, O.R., Marsden, B.G., 2004, Icarus, 170, 463

\bibitem[]{184} Milani, G.A. et al., 2007, Icarus, 187, 276

\bibitem[]{186} Mukai, T. et al., 2001, in: Interplanetary dust (ed. Gr\"unn, E., Gustafson, B.A., Dermott, S.F., Fechtig, H.), Springer, Heidelberg, p. 445

\bibitem[]{188} Rousselot, P., 2008, A\&{}A, in press

\bibitem[]{190} Sarugaku, Y. et al., 2007, PASJ, 59, L25

\bibitem[]{192} Sekanina, Z., Hanner, M. S., Jessberger, E. K., Fomenkova, M. N., 2001, in: Interplanetary dust, (ed. Gr\"unn, E.,  Gustafson, B.A., Dermott, S.F., Fechtig, H.), Springer, Heidelberg, p. 90

\bibitem[]{194} Snodgrass, C., Lowry, S.C., Fitzsimmons, A., 2006, MNRAS, 373, 1590

\bibitem[]{196} Snodgrass, C., Lowry, S.C., Fitzsimmons, A., 2008, MNRAS, accepted, arXiv:0712.4204

\bibitem[]{198} Szab\'o, Gy.M., Cs\'ak, B., S\'arneczky, K., Kiss, L.L., 2001, A\&{}A, 374, 712

\bibitem[]{200} Szab\'o, Gy.M., Kiss, L.L., S\'arneczky, K., Szil\'adi, K., 2002, A\&{}A, 384, 702

\bibitem[]{202} Weaver, H.A. et al., 1997, Science, 275, 1900 

\bibitem[]{204} Weaver, H.A. et al., 1999, Icarus, 141, 1

\end{thebibliography}
\end{document}